**Magnetic behavior of bulk and fine particles of $RCr_2Si_2C$ (R= La, Ce) compounds: Possible magnetic ordering from Cr**

K. Mukherjee, Kartik K Iyer and E.V. Sampathkumaran[*]

*Tata Institute of Fundamental Research, Homi Bhabha Road, Colaba, Mumbai 400005, India*

**Abstract**
The magnetic behavior of the quarternary compounds, $RCr_2Si_2C$ (R=La, Ce), has been investigated by magnetization *(M)* and heat-capacity *(C)* measurements (1.8 – 300 K) in the bulk polycrystals and nano forms (<1 μm) obtained by high-energy balling. The finding of emphasis is that Cr appears to exhibit magnetic ordering of an *itinerant type* at low temperatures (<20 K) in the bulk form, as inferred from a combined look of all the data. The magnetic ordering gets gradually suppressed with increasing milling time. Evidence for mixed-valence state of Ce for the bulk form is obtained from the tendency of magnetic susceptibility to exhibit a maximum above 300 K. However, this feature vanishes in the nano form, which exhibits a Curie-Weiss behavior above 200 K as though Ce tends towards trivalency in these fine particles; in addition, there is a weak upturn in *C/T* below 10 K in the bulk, which becomes very prominent in the milled Ce-based specimens at lower temperatures, as though heavy-fermion behavior gets stronger in smaller particles.

1.  **Introduction**

The ternary compounds of the form R$X_2Y_2$ (R = rare earths, $X$ = transition metals, $Y$ = Si, Ge), derived from ThCr$_2$Si$_2$-type layered tetragonal structure (space group: *I*4/*mmm*) have been attracting a lot of attention for the past four decades, as a number of phenomena due to strong electron correlations have been discovered in this family. The ease with which many 3d, 4d and 5d transitions metal ions for $X$ can be substituted and solid solutions can be formed enabled the community to address various phenomena in different ways. It should however be noted that the transition metal ions were believed not to carry magnetic moment, except for $X$= Mn [1]. This assumption was based on the absence of knowledge on the magnetic behavior of Cr-based systems for a long time, as these compounds were surprisingly subjected to extensive magnetic investigations relatively recently. The ThCr$_2$Si$_2$-type structure was found to be stable for rare earth ions from Nd to Lu only [2-5] in this Cr-series. The antiferromagnetic ordering ($T_N$ > 600 K) of Cr sublattice has been reported for TbCr$_2$Si$_2$, HoCr$_2$Si$_2$ and ErCr$_2$Si$_2$ [5]. Interestingly, CeCr$_2$Si$_2$ does not form a stable phase [6]. Nevertheless, recent reports [6 - 9] established that insertion of carbon stabilizes RCr$_2$Si$_2$C compounds with a filled CeMg$_2$Si$_2$-type structure (space group: *P*4/*mmm*) and very little work has been reported on these compounds. Among the two recent magnetic investigations [8, 9] on these new carbides, there is a disagreement with respect to the magnetism on Cr. Janatova et al [8] claim that there is a magnetic ordering below about 30 K, whereas Klosek et al [9] do not attribute any magnetic ordering to Cr, despite the fact that the plots of magnetic susceptibility ($\chi$) versus temperature ($T$) at low temperatures look similar in both the reports, for instance, for La and Ce cases. Klosek et al [9] attribute this low temperature feature to CeC$_2$ impurity present in their Ce sample.

Keeping above controversy in mind, we have reinvestigated the compounds, LaCr$_2$Si$_2$C and CeCr$_2$Si$_2$C. These two compounds have been chosen for this purpose as there is no magnetic ordering from the rare-earth ions (including Ce as it is mixed-valent [8,9]). In addition, we have investigated the magnetic behavior of these compounds in the fine particle form (less than 1 micron) as a part of our recent initiative [10] to probe electron correlation effects in Ce systems in nano form. The main conclusions are: (i) One can not rule out itinerant magnetism from Cr for the bulk form, supporting Janatova et al [8], (ii) possible magnetism from Cr appears to get gradually suppressed with a reduction in particle size, and (iii) Ce is distinctly mixed-valent in the bulk form and the fine particles of Ce are characterized by higher values of $\chi$ in the paramagnetic state which could imply that the magnetic moment on Ce is influenced with varying particle size.

2.  **Experimental details**

The polycrystalline samples of LaCr$_2$Si$_2$C and CeCr$_2$Si$_2$C were synthesized by repeated arc melting of stoichiometric amounts of constituent elements (purity > 99.9% for R, and >99.99% for the rest) in an atmosphere of argon. The samples were re-melted several times to improve the homogeneity. X-ray diffraction patterns (XRD) (Cu K$_\alpha$) (figure 1) confirm the formation of proper phases. Though there are a few weak unidentified peaks for La case, we could not find any extra line in the case of Ce sample, in contrast to Ref. 9. The lattice constants obtained by Rietveld refinement are found to be $a$= 4.063(4) Å, $c$= 5.430 Å for La compound, and $a$= 4.019(4) Å, $c$= 5.290 Å for Ce compound. Back-scattered images from field-emission scanning electron microscopic (FE-SEM) further confirmed phase purity. The materials thus obtained (called L_B for LaCr$_2$Si$_2$C and C_B for CeCr$_2$Si$_2$C) were milled in a medium of toluene for 3 hrs (called L_N1 for LaCr$_2$Si$_2$C and C_N1 for CeCr$_2$Si$_2$C) and then for another 9 hrs (called L_N2 for LaCr$_2$Si$_2$C and C_N2 for CeCr$_2$Si$_2$C) in a planetary ball bill (Fritsch pulverisette-7 premium line) operating at a speed of 500 rpm. Zirconia vials and balls of 5mm diameter (without any magnetic impurity) were used and the ball-to-material ratio was kept at 10:1. The phase stability without any change in crystal structure after ball-milling was ascertained by x-ray diffraction (see figure 1). It is to be noted that the intensities of the weak extra lines observed for L_B get reduced with milling. Rietveld refinement of the diffraction patterns show that the changes in the lattice parameters *a* and *c* are, respectively, less than 0.09% and 0.2% for L_N1 and L_N2, while it is less than 0.2% and 0.3% for C_N1 and C_N2 with respect to bulk forms. This implies that a reduction in particle size introduces insignificant changes in structural parameters as defined by peak positions, but the base of the diffraction lines apparently reveal some broadening. Since the shape of the background of the XRD patterns are not found to be altered by milling (as inferred from reasonable values of R-factors as well in Reitveld fitting), we infer that there is no noticeable amorphization. Since strains also can contribute to XRD line-broadening, we resorted to FE-SEM to infer about a reduction in particle size. The FE-SEM images on milled specimens are show in figure 2. A careful look at these images reveals that the particles seem to be significantly agglomerated, but the observed sizes of the majority of particles were much smaller than 500 nm after milling for 12 hrs (as inferred from probing different regions of the sample). Williamson-Hall plots (usually employed to obtain particle size from x-ray diffraction data) were not found to be linear possibly due to significant differences in the shapes of particles. Hence an average value from Debye-Scherrer formula was estimated from the x-ray diffraction lines in the range $2\theta$= 28–45 degree; this value falls in the range ~ 60 Å, 30 Å, 100 Å and 30 Å for LN1, LN2, CN1 and CN2 respectively without accounting for strains. Since there appears to be a large distribution of particle size as revealed by SEM images, we tend to use the phrase 'fine particles' in this article. The dc magnetization [*M(T)*: (1.8 – 300 K); and *M(H)*: up to *H*= 50 kOe)] measurements were performed with the help of a commercial SQUID magnetometer (Quantum Design) and heat-capacity (C) measurements in zero-field as well as in 50 kOe were carried out with a commercial physical properties measurements system (Quantum Design); we used a very small amount of GE varnish as the binding material for a known mass of fine particles, as the heat-capacity values of this varnish were found to be negligible compared to those of samples in the temperature range of interest.

3. **Results and discussion**
3.1. **Magnetic behavior of bulk specimens**

In figure 3, we plot magnetization as a function of temperature below 60 K for the bulk specimens, measured in the presence of two magnetic fields, 100 Oe and 1 kOe. For *H*= 100 Oe, the curves obtained in zero-field-cooled (ZFC) and field-cooled (FC) conditions (from 60K) of the alloys tend to bifurcate at low temperatures. For *H*= 1 kOe, these curves are indistinguishable. The foremost conclusion is that the variation of *M* with *T* is rather weak as the *T* is lowered in the range 30-60 K, but there is a sudden upturn at lower temperatures, as though there is a magnetic transition (also visible in figure 3e). The temperature derivative of $\chi$ (measured in 100 Oe) exhibits a peak near 20 and 14 K for La and Ce samples, and at nearly same temperatures, ZFC-FC curves bifurcate; therefore, these temperatures can be defined as the magnetic ordering temperature (see insets of figure 3a and 3b). Since the ZFC and FC curves monotonically vary without any peak in ZFC-curve, one can assume that there is no spin-glass freezing; in addition, we did not find any frequency dependence in ac $\chi$ (not shown here), supporting this conclusion. At this juncture, it should be mentioned that such a low-temperature upturn in χ was reported by two groups, Janatova et al [8] and Klosek et al [9], though the latter attribute it to $CeC_2$ impurity phase. It should be emphasized that, unlike in Ref. 9, we do not find any extra peaks in x-ray diffraction pattern that could be attributable to this impurity phase; in addition, the upturn observed in the La sample makes an argument in terms of $CeC_2$-phase untenable. *In addition, the magnitude of the upturn (say, the value of M at 1.8 K for a given field) in our case without $CeC_2$ is comparable to that in figure 6 of Ref. 9 (with extra lines in x-ray diffraction pattern), which is not possible to explain in terms of any impurity effect.* We therefore propose that the observed magnetic ordering is intrinsic to Cr in these compounds.

In order to get more insight into the nature of magnetic ordering, we have measured isothermal magnetization at 1.8 K (see figures 3c). Following a sharp increase for an initial application of *H* typical of ferromagnets, for instance, in the case of La compound, there is a gradual increase of *M* varying nonlinearly with *H*. The magnitude of the magnetic moment at high fields is very small (of the order of 0.08 $\mu_B$ per formula unit at 50 kOe) and such a small fractional value is a characteristic feature of itinerant magnetism. It may be stated that remnant magnetization is negligible. These observations are in good agreement with Ref. 8. Similar observations are made in the case of Ce compound, though a weak hysteresis is observed in this case.

We have also measured heat-capacity as a function of temperature below 50 K and we did not any λ-anomaly and therefore it appears that the entropy associated with the magnetic transition is negligible, which is consistent with itinerant magnetism. Such a very low moment associated with itinerant magnetism could be the reason why it escaped detection in neutron diffraction experiments [9]. An important observation we have made in *C/T* data plot in figure 3d is that, following a linear variation with decreasing temperature below 20 K, there is an upturn at low temperatures below about 10 K for the Ce case, which is absent in the case of La compound. This upturn is similar to that known for many Ce-based heavy fermion systems, e.g., $CeCu_2Si_2$ [Ref. 11], though it is surprising that an application of a magnetic field as large as 50 kOe does not depress this behavior.

Finally, in the paramagnetic state of Ce compound, we find a distinct evidence for increasing χ with increasing temperature (figure 3e, inset). This is a well-known characteristic feature of intermediate valence systems of Ce with Kondo temperature above 300 K. In the case of La compound, there appears to be a Curie-Weiss behavior in the paramagnetic state; the temperature dependence of the susceptibility at 5 kOe has to be fitted with modified Curie Weiss law ($\chi = \chi_0 + C/(T-\theta_p)$, where $\chi_0$, C and $\theta_p$ carry usual meanings) above 200 K for L_B sample, as the magnitude of the temperature independent component ($\chi_0$) is comparable to the Curie-Weiss term; the obtained effective magnetic moment value is around 0.9 $\mu_B$/formula unit (attributable to Cr), which is about two orders of magnitude larger than that observed at high fields in the magnetically ordered state – which is a characteristic feature of itinerant magnetism.

## 3.2 Magnetic behavior of fine particles

Now we focus on the magnetic behavior of the fine particles of the above compounds. From figure 4, it is seen that, with increasing milling time (that is, with decreasing average particle size), χ decreases as one moves from L_B to L_N1 to L_N2, while the reverse trend is interestingly observed for Ce compound (figure 4b). This is very transparent from the plot of inverse χ versus *T* obtained in a field of 5 kOe, shown in the insets of figure 4. In fact, increasing χ with increasing *T* seen in the bulk form for Ce near 300 K is absent in the fine particle form and this is replaced by a Curie-Weiss regime (seen above 200 K).   This behavior in contrast to that noted for La specimens in the paramagnetic state has to be attributed to an increase in the magnetic moment of Ce with decreasing particle size, naturally due to an increase in the mean valence of Ce. Possible role of surface-induced changes in the valence Ce as a cause of this increase can not be ruled out. A distribution in the average valence of the Ce ions due to a distribution in particle size could also contribute to the broadened base of the x-ray diffraction peaks.  From the Curie-Weiss regime (200-300 K), we derive the magnetic moment which is about 2.9 $\mu_B$ per formula unit, which is slightly larger than that expected  for trivalent Ce ion, and its origin can be attributed to a contribution from Cr.

At low temperatures, the bifurcation of ZFC-FC $\chi$ curves (obtained in a field of 100 Oe) and the peak in $d\chi/dT$ versus T plots shift to lower temperatures with the decrease in particle size for both the compounds (see inset of figures 4a and 4b). The bifurcation of these curves is interestingly visible even for $H$= 1 kOe for all except in the case of C_N2. For the C_N2 sample, the ordering temperature, as defined by the peak temperature in $d\chi/dT$,  if it exists, must be below 1.8 K and hence we do not see ZFC-FC  bifurcation above 1.8 K. It is thus obvious that the magnetic behavior of La and Ce compounds in fine particles differ in a subtle manner, possibly due to changes in Ce valence which in turn influences Cr magnetism through changes in 4f hybridization and electronic structure.

The contrasting behavior of $\chi$ versus $T$ curves of fine particles of La and Ce compounds is also reflected in the $M(H)$ isotherms at 1.8 K (figures 5a and 5b). It is distinctly clear from figure 5a that the magnetization value at a given field is lower for the fine particles of the La case, while it increases dramatically for the Ce case; this supports the conclusion that there is an increase in Ce valence. Otherwise, the nature of the $M$ versus $H$ curves is qualitatively the same as that in respective bulk forms. The magnetization value at 1.8K and 50 kOe for C_N2 is 0.19$\mu_B$, which is ~ 20% of the theoretical value of ferromagnetically ordered Ce ions with doublet ground state. No further inference could be drawn from these data, as $M$ at high fields is far from saturation.

With respect to heat-capacity behavior (figure 6), like in bulk form, the anomalies due to magnetic ordering are absent in the fine particles. In the case of L_N2, for instance, the $C/T$ versus  $T^2$ plot is essentially linear in the range 5-10 K, and the value of the linear term, $\gamma$ (obtained from   $C = \gamma T + \beta T^2$) is ~ 33 mJ/mol K$^2$. The value of $\gamma$ is quite large, possibly attributable to Cr itinerant magnetism. It is interesting that magnetic field has no effect on this $\gamma$ value. There is a drop at lower temperatures, which could signal some interesting effects due to Cr magnetism.

The cerium containing fine particles exhibit significant low temperature anomalies in $C/T$ plots, which are different from those seen for La cases. The upturn seen in $C/T$ versus  $T$ below 10 K  for C_B sample is suppressed to  a low temperature (<5 K) and the temperature at which the upturn occurs increases marginally with magnetic field, as demonstrated for 20 and 50 kOe for C_N2. Though there is a minimum in $C/T$ as in bulk form, the magnitude of the upturn gets more significant in fine particles. The value of $C/T$ (in zero field) increases from about ~ 20 mJ/mol K$^2$ at about 10 K to ~ 32 mJ/mol K$^2$ at 2 K for $C\_B$ sample, whereas the corresponding upturn setting in for C_N2 at ~ 110 mJ/mol K$^2$ (at about 5 K) rises to ~ 200 mJ/mol K$^2$ (at 2 K). This implies strengthening of heavy-fermion character with decreasing particle size, induced by the tendency towards localization of Ce 4f electrons.

## 4. Summary

We have synthesized C-filled $LaCr_2Si_2$ and $CeCr_2Si_2$ and studied their magnetic characteristics. Though we could not detect $CeC_2$ or any other magnetic impurity, we find distinct magnetization anomalies near 20 K in both the compounds. Viewed together with heat-capacity, effective moment and isothermal magnetization data, we infer that Cr could exhibit itinerant magnetism in these compounds. The magnetic ordering temperature of Cr appears to undergo a decrease with a reduction in particle size (to <1 micron). There is a clear evidence for strong mixed-valence of Ce in the bulk form of $CeCr_2Si_2C$ and the 4f electrons get more localised in the fine particles, giving rise to enhanced heavy-fermion behavior at low temperatures.

**Acknowledgements**
We are thankful to N.R. Selvi and G.U. Kulkarni, Jawaharlal Nehru Center for Advanced Scientific Research, Bangalore, India, for SEM measurements.

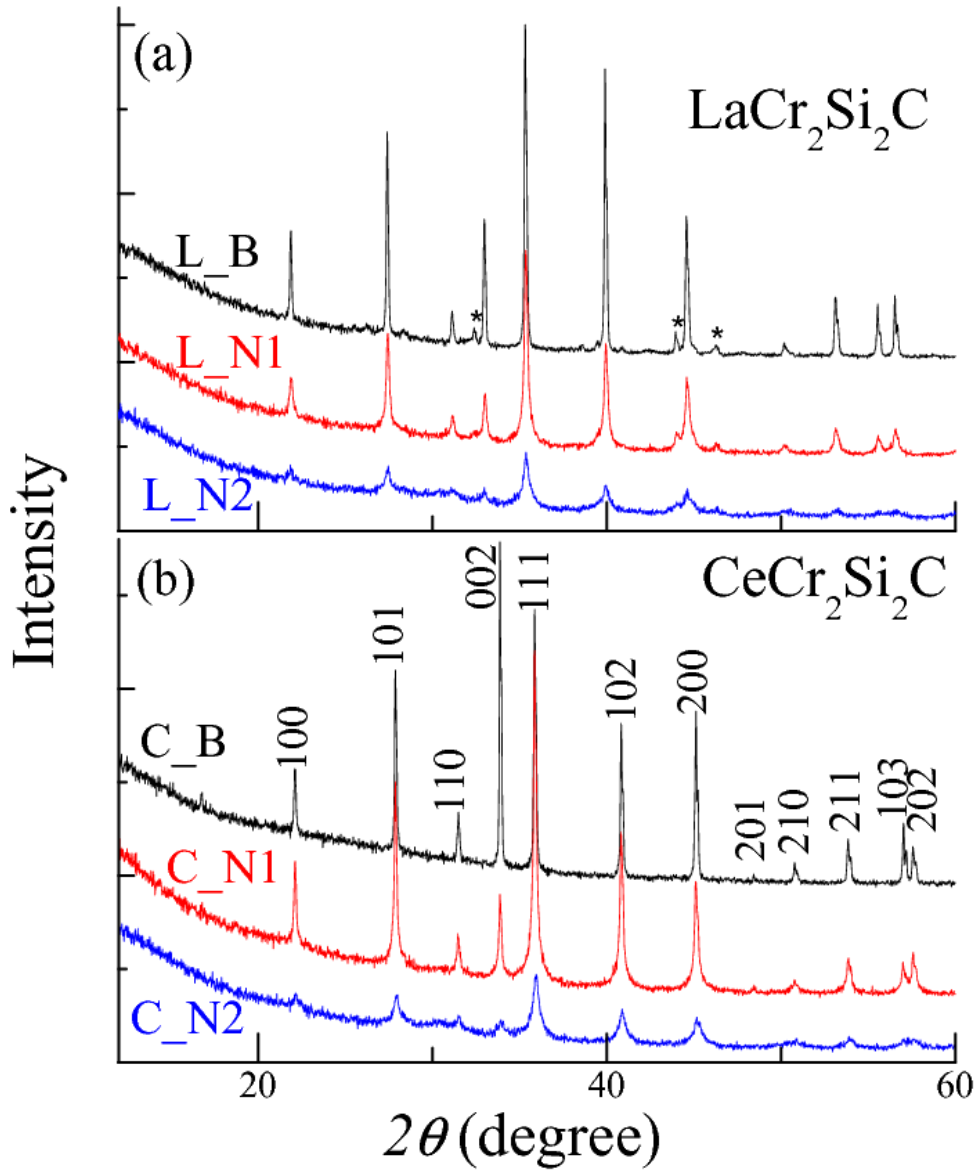

Figure 1:
(color online) X-ray diffraction patterns (Cu $K_\alpha$) for the bulk and the milled specimens of: a) $LaCr_2Si_2C$ and b) $CeCr_2Si_2C$. In one case, the miller indices are given. The asterisks mark unidentified lines.

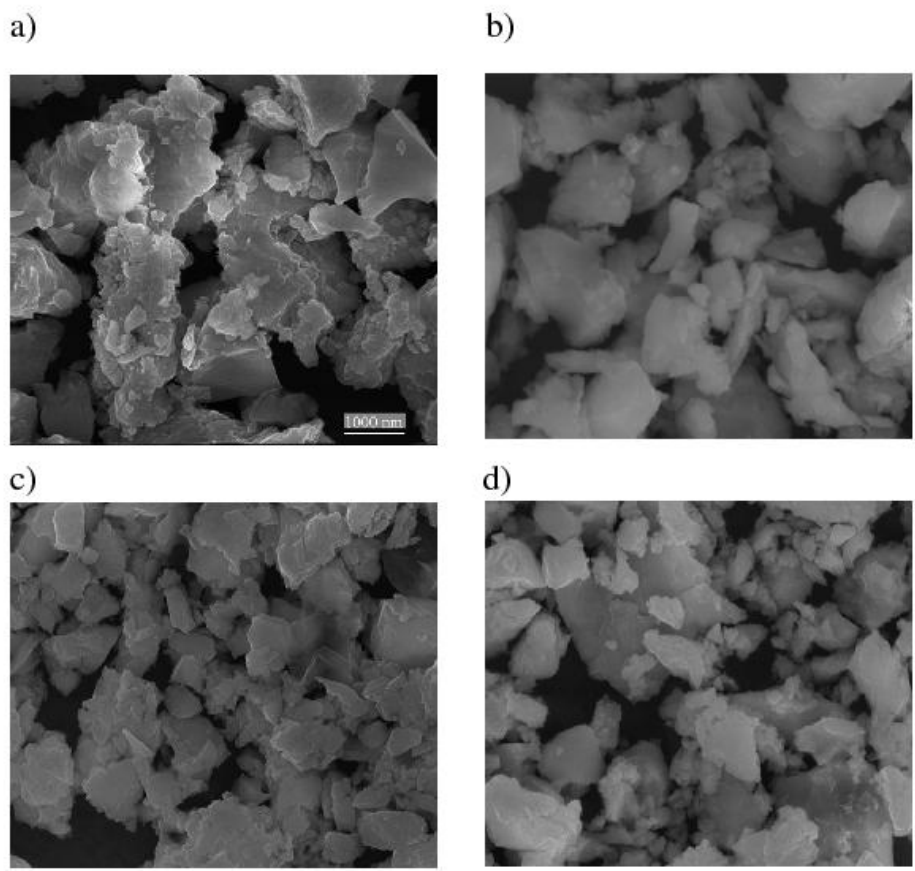

Figure 2:
(color online) Scanning electron microscopic images of the milled specimens of LaCr$_2$Si$_2$C and CeCr$_2$Si$_2$C.: **(a)** for L_N1; **(b)** for L_N2; **(c)** for C_N1 and **(d)** for C_N2. The scale mentioned in (a) is the same for all figures.

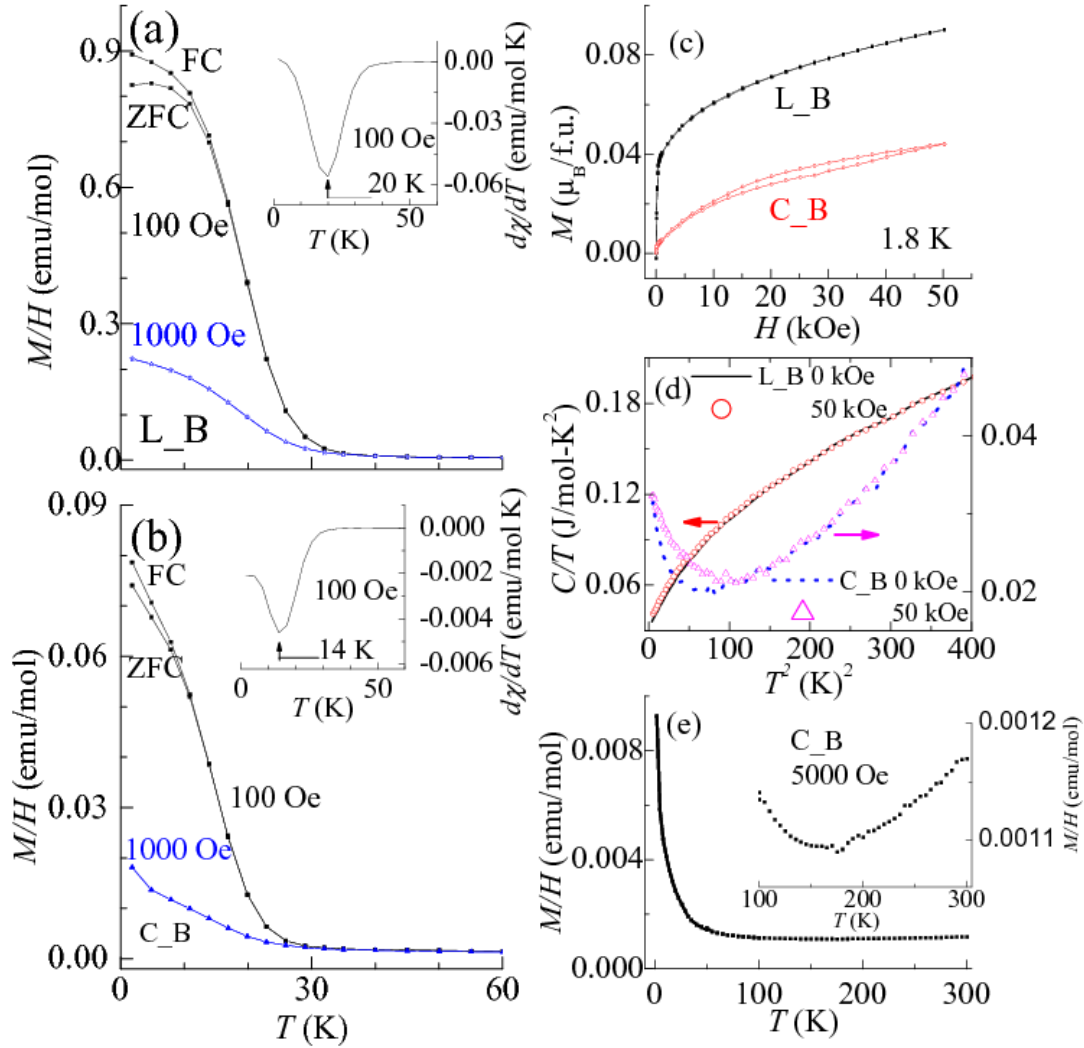

Figure 3:
(color online) Temperature ($T$) dependence (below 60 K) of magnetization ($M$) divided by magnetic field ($H$) for the bulk specimens of **(a)** L_B of LaCr$_2$Si$_2$C and **(b)** C_B of CeCr$_2$Si$_2$C obtained in a magnetic field of 100 Oe and 1 kOe; in the case of 100 Oe, the bifurcation of the curves obtained for field-cooled and zero-field-cooled conditions is shown. In the insets, the temperature-derivative of magnetic susceptibility ($\chi$) as a function of $T$ is shown. In **(c)** Isothermal magnetization at 1.8 K, **(d)** heat-capacity divided by $T$ versus $T^2$ in zero-field and in 50 kOe for LaCr$_2$Si$_2$C and CeCr$_2$Si$_2$C, and **(e)** $M/H$ as a function of temperature for CeCr$_2$Si$_2$C measured in a field of 5 kOe are plotted and the data above 100 K is plotted in an expanded scale in the inset. The lines through the data points serves as guides to the eyes.

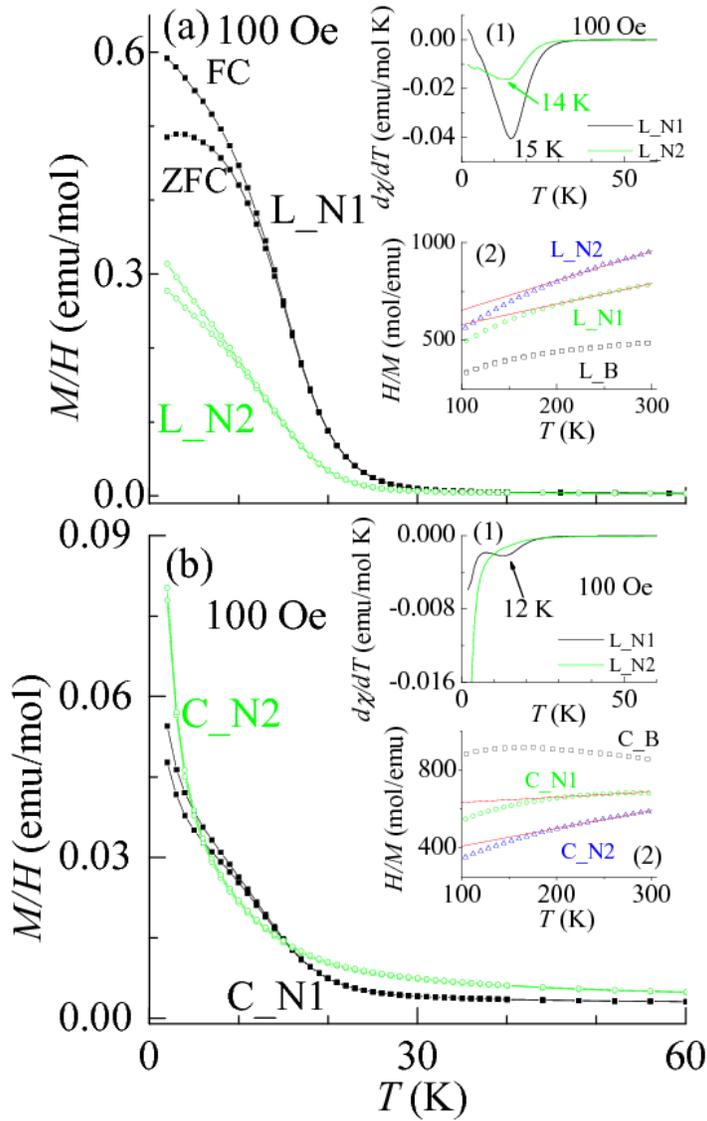

Figure 4:
(color online) Temperature (*T*) dependence (below 60 K) of magnetization (for both zero-field-cooled and field-cooled conditions) divided by magnetic field obtained in 100 Oe for the specimens **(a)** L_N1 and L_N2 and **(b)** C_N1 and C_N2. The ZFC-FC curves overlap for C_N2. The lines through the data points serve as guides to the eyes. In the insets, the *T*-derivative of *M/H* measured in 100 Oe and *H/M* measured in 5 kOe are plotted as a function of *T* (along with the curves for bulk specimens) are plotted; in the latter graphs, the continuous lines represent Curie-Weiss regime.

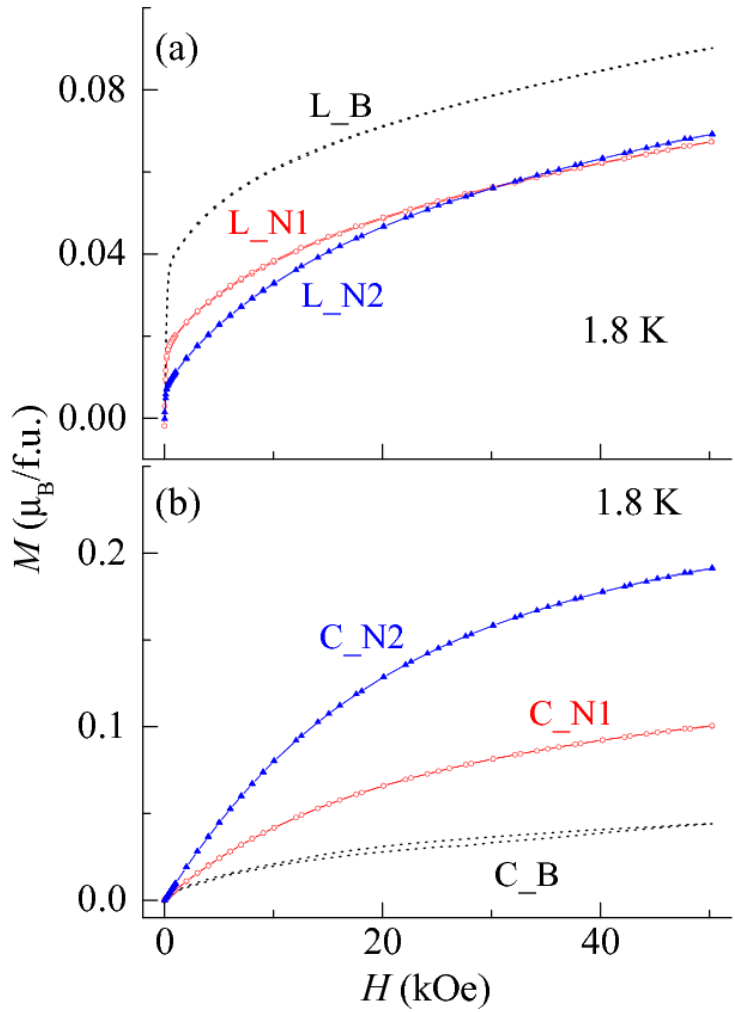

Figure 5:
(color online) Isothermal magnetization at 1.8 K for all milled specimens of **(a)** $LaCr_2Si_2C$ and **(b)** $CeCr_2Si_2C$. For comparative purposes, the data for the bulk specimens are also included. The lines through the data points serve as guides to the eyes.

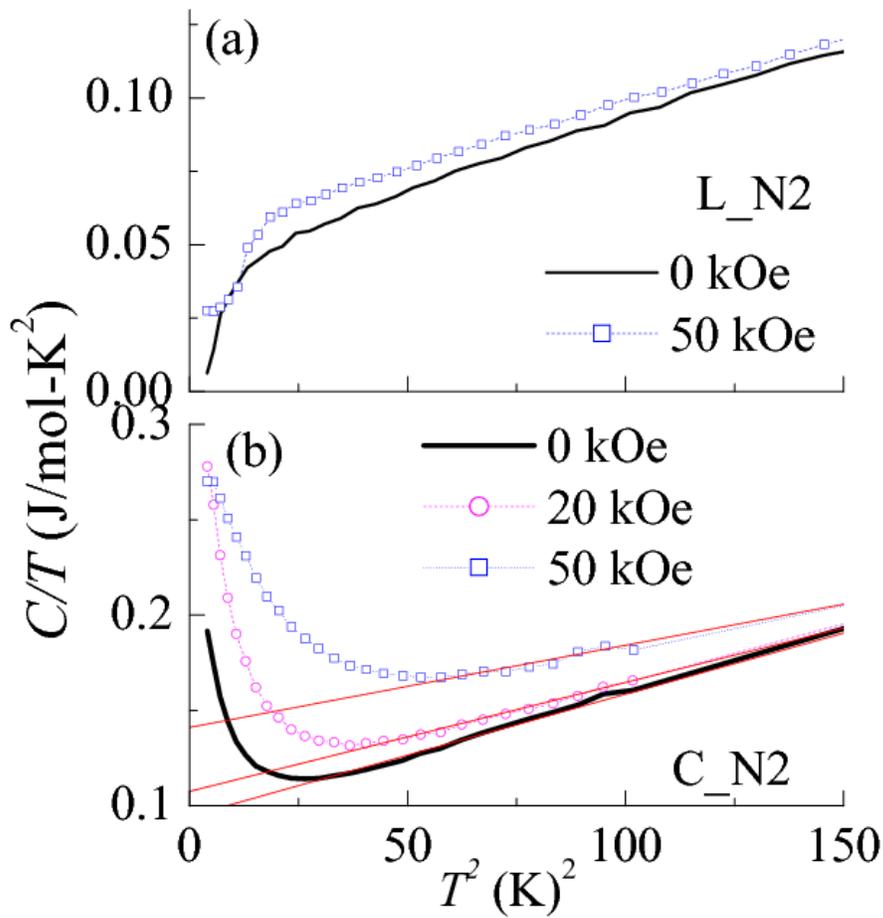

Figure 6:
(color online) Heat capacity divided by temperature versus square of temperature for the specimens **(a)** L_N2 of $LaCr_2Si_2C$ and **(b)** C_N2 of $CeCr_2Si_2C$ in the presence of various fields.